\documentstyle[11pt,paspconf,psfig,epsf]{article}

\newcommand{\OA}{\mbox{$A$}}
\newcommand{\OB}{\mbox{$B$}}
\newcommand{\OAR}{\mbox{$A(R)$}}
\newcommand{\OBR}{\mbox{$B(R)$}}

\newcommand{\RSUN}{\mbox{$R_0 \;$}}
\newcommand{\RSUNn}{\mbox{$R_0$}}
\newcommand{\VSUN}{\mbox{$\Theta_0 \;$}}
\newcommand{\VSUNn}{\mbox{$\Theta_0$}}

\newcommand{\pmt}{\mbox{$\pm \;$}}
\newcommand{\rtp}[1]{\mbox{$^{#1}$}}
\newcommand{\kms}{\mbox{${\rm km \;s}^{-1}$}}

\newcommand{\mVskip}[1] {\vspace*{ #1}}

\newcommand{\ssRO}{\renewcommand{\baselinestretch}{0.85}\begin{small}}
\newcommand{\esRO}{\end{small}\renewcommand{\baselinestretch}{1.0}}

\def\edcomment#1{\iffalse\marginpar{\raggedright\sl#1\/}\else\relax\fi}
\reversemarginpar

\begin{document}

\title{Refining the Oort Constants: the case for a smaller Milky Way}

\author{Rob P. Olling, Michael R. Merrifield}
\affil{University of Southampton, Department of
Physics and Astronomy, Southampton SO17 1BJ, United Kingdom}

\author{{\em to appear in the proceedings of the
   "Workshop on Galactic Halos'',
   Santa Cruz, August 1997 (ASP Conference Series}}

\author{}
\affil{}

\vspace*{-15mm}
\begin{abstract}

The local stellar kinematics of the Milky Way, parameterized by the Oort
constants $A$ and $B$, depend on the local gradient of the rotation
curve, its absolute value ($\Theta_0$), and the distance to the Galactic
center ($R_0$).  The surface density of interstellar gas in the Milky
Way varies non-monotonically with radius, and so contributes
significantly to the local gradient of the rotation curve, and the Oort
constants.  Because of this, the Oort {\em functions} \OAR\ and \OBR\
differ significantly from the dominant $\sim \Theta_0/R$ dependence, in
the Solar neighborhood and other locations in the Galaxy.  These models
may explain the $\sim$40\% difference between the values for $2 A \RSUN$
derived from radial velocity data originating in the inner and outer
Galaxy (Merrifield 1992).  Incorporating these local non-linearities
explains the significant differences between the Oort constants derived
from nearby stars ($d \le 1$ kpc; Hanson 1987=H87) and distant Cepheids
($d= 0.5-6$ kpc; Feast \& Whitelock 1997=FW97).  However, a consistent
picture only emerges if one adopts small values for the Galactic
constants: \RSUN = 7.1 \pmt 0.4 kpc, and \VSUN = 184 \pmt 8 \kms.  These
values are consistent with most kinematical methods of determining
\RSUNn, including the proper motion of Sgr A$^*$ (Backer \ 1996), the
direct determination of \RSUN using water masers (7.2 \pmt 0.7 kpc, Reid
1993), and constraints set by the shape of the Milky Way's dark halo
(Olling \& Merrifield 1997b=OM97b). 

\end{abstract}

\mVskip{-8mm}
\section{Introduction}
\mVskip{-2mm}

Due to our location within the Milky Way and the modest uncertainties in
\RSUN (7.7 \pmt 0.7 kpc; Reid 1993) and \VSUN (200 \pmt 20 \kms; Sackett
1997), the rotation curve of the Milky Way, $\Theta(R)$, is difficult to
establish (Fich \& Tremaine 1991; Olling \& Merrifield 1997a=OM97a). 
Stellar kinematical
data in the form of proper motions and radial velocities can be used to
constrain the Galactic constants via the Oort {\em functions}:
$A(R)=\frac{1}{2} \left( \frac{\Theta(R)}{R} - \frac{d\Theta(R)}{dR}
\right)$ and $B(R)=-\frac{1}{2} \left( \frac{\Theta(R)}{R} +
\frac{d\Theta(R)}{dR} \right)$.

Unfortunately, the available observations of the Milky Way's rotation
curve are not good enough to calculate the derivatives of $\Theta(R)$
directly.  Instead, we fit mass models to the observations and calculate
the derivatives from the model rotation curves.  The dominant
contributors to the total mass are the stellar disk and the dark matter
(DM) halo, which are believed to be fairly smoothly distributed with
radius.  However, the distribution of interstellar hydrogen (ISM) shows
density enhancements such as rings and arms, which will produce a
contribution to $\Theta(R)$ that varies non-monotonically with radius. 
This effect gives rise to local features superimposed on the dominant
$\Theta/R$ behavior of the Oort functions (Fig.~1).  On larger scales,
the Oort functions follow the no-ISM relations (dotted line). 

\begin{center}
 
\psfig{file=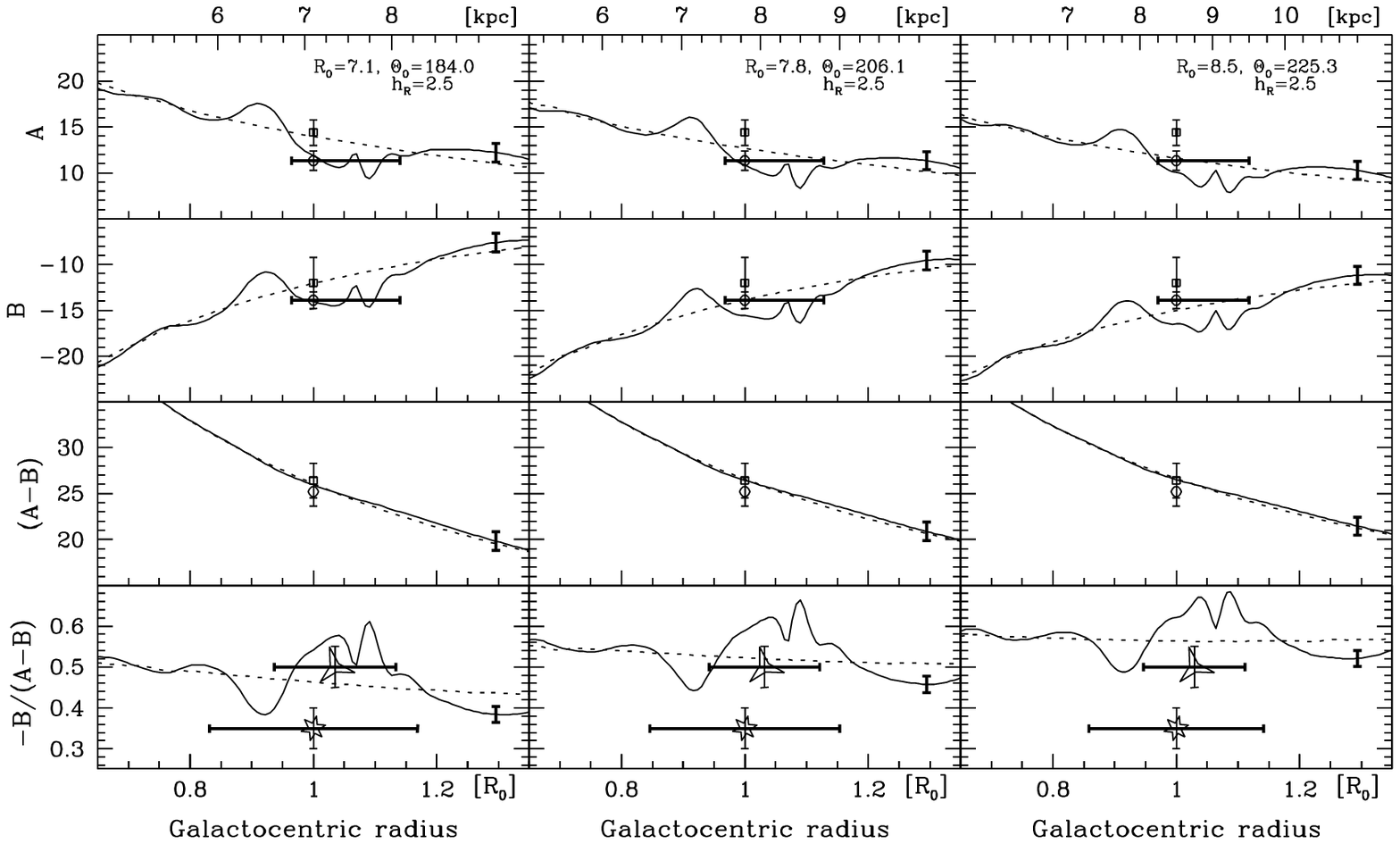,height=7cm,width=13cm}

\ssRO

\noindent \parbox{13cm}{ Fig.~1.  The Oort functions, \OAR\, \OBR, and
$(A-B)$ [\kms kpc\rtp{-1}] as derived for three model rotation curves. 
The solid lines are derived from the full mass model, the dashed lines
have no gas component.  The various observational estimates of these
quantities are also shown, with the horizontal error bars indicating the
radial range over which the observations effectively averaged.  H87's
and the IAU standard values (Kerr \& Lynden-Bell 1986=KLB86) are plotted
(squares and circles), as well as the velocity dispersion ratios of
local stars (triangles and hexagons, bottom panel).  Notice the
$\sim$40\% difference between the values of $2AR_0$ inferred from
extrapolating the inner and outer Galaxy data, similar to Merrifield's
(1992) observational findings. 
}

\esRO

\end{center}

\vspace{-0.05em}
   From Fig.~1 it is clear that if the radial extend of the stellar
kinematical surveys is more than a few hundred parsec, it is imperative
to take the slope in the Oort functions (a few \kms kpc\rtp{-2}) into
account.  However, notice that \OAR\ and \OBR\ are almost flat in the
first kpc beyond the Solar circle.  This is the region sampled by the
Lick Northern Proper Motion stars used in Hanson's (1987) determination
of the Oort constants.  Thus, we compared his values (\OA=11.3 \pmt 1.1,
and \OB=-13.9 \pmt 0.9 \kms kpc\rtp{-1}) with our model predictions.  We
also use the combinations ($A-B$) and $-B/(A-B)$ as constraints (for
details, see OM97a).  Inspection of Figure~1, reveals that models with
small values for the Galactic constants fit the observations better than
the values currently considered best (middle panels) and the IAU values
(rightmost panels). 

\mVskip{-3mm}
\section{Results}
\mVskip{-3mm}

   We can formalize the constraints placed on the values of the Galactic
constants by calculating a $\chi^2$ statistic comparing the five
observed combinations of the Oort constants in Fig.~1 to the values
predicted by the models.  Because of the radial dependence of the Oort
functions, we compared the model and observed values over the radial
extend of the observations (horizontal error bars).  Since these regions
are approximately equal to the size of the epicycles of the stellar
populations studied, we expect that the Oort functions can show
structure on these scales.  The $\chi^2$ statistics were calculated for
a range of values for $R_0$ and $\Theta_0$.  The best-fit
(minimum-$\chi^2$) values are: $R_0 = 7.1 \pm 0.4$ kpc, and $\Theta_0 =
184 \pm 8$ \kms.  In Figure~2, we plot the probability that any given
values for $R_0$ and $\Theta_0$ are consistent with the observed Oort
constraints.  For example, the official IAU-sanctioned values of $R_0 =
8.5$ kpc and $\Theta_0 = 220$ \kms are ruled out at the 99\% confidence
level.  Comparing our best fit values with \RSUN determinations based on
kinematical constraints (see Reid 1993 for a compilation), we find that
all are consistent with the leaner Galaxy we propose here.  In
particular, \RSUNn=7.1 kpc is entirely consistent with its only {\em
direct} determination employing H$_2$O masers proper motions (\RSUNn=
7.2 \pmt 0.7 kpc, Reid 1993).  Furthermore, these Galactic constants are
consistent with the proper motion of Sgr A$^*$ (Backer 1996).  From a
new and completely independent analysis based on the shape of the
Galaxy's dark matter halo, we find similar constraints on the Galactic
constants (OM97b). 

\psfig{file=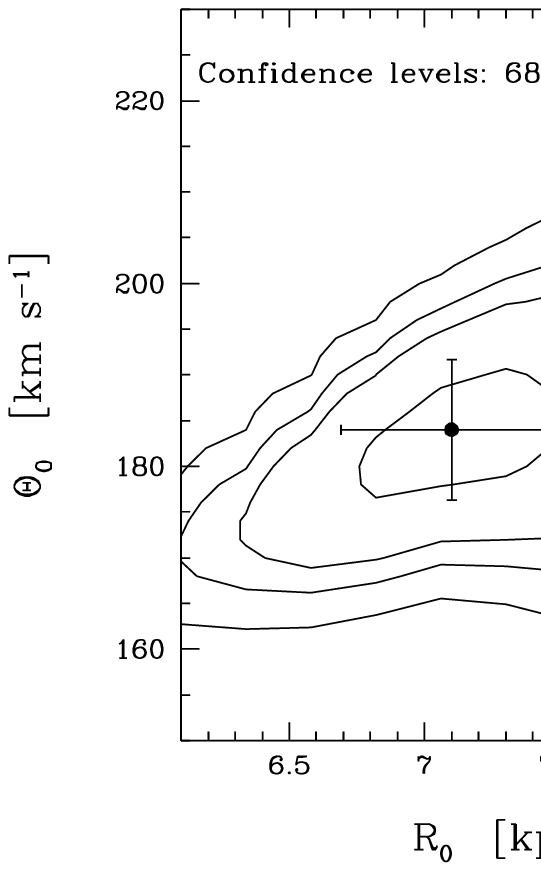,height=5cm,width=6cm}

\ssRO

\vspace*{-4.5cm} \noindent \parbox{6cm}{ Fig.~2.  The contours of equal
likelihood as a function of \RSUN and \VSUNn, calculated by comparing
the model values with the observed $A$, $B$, $A-B$, and $-B/(A-B)$
constraints presented in Fig.~1.  The best-fit minimum-$\chi^2$ values
for \RSUN, \VSUNn, and their 1-$\sigma$ errors are also indicated.  The
IAU standard values are \RSUNn=8.5 kpc, and \VSUNn=220 \kms.}

\esRO

\vspace{15mm}

The Oort constants derived from nearby stars ($d\le 1$ kpc, H87) differ
significantly ($\sim$3$\sigma$) from the values derived at large
distances ($d = 0.5 - 6$ kpc, FW97).  However, extrapolating our best
fit model from the distant Galaxy towards the Solar position, i.e.,
following the no-ISM line in Fig.~1, yields \OA\ and \OB's very close to
those of FW97.  Furthermore, our models and the FW97 models predict
almost identical Cepheid proper motions.  We conclude that the
discrepancy between the H87 and FW97 Oort constants is caused by the
non-linear behavior in the Solar neighborhood and that a consistent
picture only emerges for a leaner Milky Way with \RSUN=7.1 \pmt 0.4 kpc,
\VSUNn=184 \pmt 8 \kms.

\ssRO

\mVskip{-3mm}

\esRO

\end{document}